# Are there any good digraph width measures? [†]

Robert Ganian[1], Petr Hliněný[1], Joachim Kneis[2], Daniel Meister[2], Jan Obdržálek[1], Peter Rossmanith[2], and Somnath Sikdar[2]

[1] Faculty of Informatics, Masaryk University, Brno, Czech Republic
{xganian1,hlineny,obdrzalek}@fi.muni.cz

[2] Theoretical Computer Science, RWTH Aachen University, Germany
{kneis,meister,rossmani,sikdar}@cs.rwth-aachen.de

**Abstract.** Several different measures for digraph width have appeared in the last few years. However, none of them shares all the "nice" properties of treewidth: First, being *algorithmically useful* i.e. admitting polynomial-time algorithms for all $\text{MSO}_1$-definable problems on digraphs of bounded width. And, second, having nice *structural properties* i.e. being monotone under taking subdigraphs and some form of arc contractions. As for the former, (undirected) $\text{MSO}_1$ seems to be the least common denominator of all reasonably expressive logical languages on digraphs that can speak about the edge/arc relation on the vertex set. The latter property is a necessary condition for a width measure to be characterizable by some version of the cops-and-robber game characterizing the ordinary treewidth. Our main result is that *any reasonable* algorithmically useful and structurally nice digraph measure cannot be substantially different from the treewidth of the underlying undirected graph. Moreover, we introduce *directed topological minors* and argue that they are the weakest useful notion of minors for digraphs.

[†] This work is supported by the Deutsche Forschungsgemeinschaft and the Czech Science Foundation.

# 1 Introduction

An intensely investigated field in algorithmic graph theory is the design of graph *width parameters* that satisfy two seemingly contradictory requirements: (1) graphs of bounded width should have a reasonably rich structure; and, (2) a large class of problems must be efficiently solvable on graphs of bounded width. For undirected graphs, research into width parameters has been extremely successful with a number of algorithmically useful measures being proposed over the years, chief among them being treewidth [16], clique-width [6], branchwidth [18] and related measures (see also [3]). Many problems that are hard on general graphs turned out to be tractable on graphs of bounded treewidth. These results were combined and generalized by Courcelle's celebrated theorem which states that a very large class of problems ($MSO_2$) is tractable on graphs of bounded treewidth [4].

However, there still do not exist *directed graph* width measures that are as successful as treewidth. This is because, despite many achievements and interesting results, most known digraph width measures do not allow for efficient algorithms for many problems. During the last decade, many digraph width measures were introduced, the prominent ones being directed treewidth [12], DAG-width [2, 14], and Kelly-width [11]. These width measures proved useful for some problems. For instance, one can obtain polynomial-time (XP to be more precise) algorithms for HAMILTONIAN PATH on digraphs of bounded directed treewidth [12] and for PARITY GAMES on digraphs of bounded DAG-width [2] and Kelly-width [11]. But there is the negative side, too. HAMILTONIAN PATH, for instance, likely cannot be solved on digraphs of directed treewidth, DAG-width, or Kelly-width at most $k$ in time $O(f(k) \cdot n^c)$, where $c$ is a constant independent of $k$. Note that HAMILTONIAN PATH *can* be solved in such a running time for undirected graphs of treewidth at most $k$ [4].

Additionally, for the newly introduced [9] DAG-depth and Kenny-width[3] – digraph width measures that are much more restrictive than DAG-width – problems such as DIRECTED DOMINATING SET, DIRECTED CUT and $k$-PATH remain NP-complete on digraphs of constant width [9]. In contrast, another recent digraph measure bi-rank-width [13] looks more promising. A Courcelle-like $MSO_1$ theorem exists for digraphs of bounded bi-rank-width, and many other interesting problems can be solved in polynomial (XP) time on these [13, 10]. For a recent survey on complexity results for DAG-width, Kelly-width, bi-rank-width, and other digraph measures, see [9].

In this paper, we boldly ask whether there exist digraph width measures that are algorithmically useful, and if so what properties can they be expected to satisfy. We first address the question of what it means for a width measure to be *algorithmically useful*. While there is no formal definition of this notion, we appeal to what is known about width measures for undirected graphs, in particular, about treewidth. As mentioned earlier, Courcelle's Theorem states that all problems expressible in $MSO_2$ logic are (fixed-parameter, or FPT) tractable on graphs of bounded treewidth. It would be nice to have such a strong result for a digraph width measure, but to this day there exists no widely accepted logical language specifically aimed at digraphs at all. This fact then prompts us to consider the *least common denominator* of all possible descriptive languages over digraphs that; a) have sufficient expressive power (meaning they can quantify over sets, not only over singletons), and b) can identify the arc/edge relation over the vertex set. Clearly, this least common denominator includes at least the ordinary $MSO_1$ logic (see Section 2) of the underlying undirected graph.

We thus define algorithmic usefulness as the property of admitting polynomial-time (XP to be precise) algorithms for all $MSO_1$-definable problems on digraphs of bounded width as the parameter. Note that we even relax the required time bound from FPT-time to XP-time (see in Section 2). It is easy to see that algorithmically useful digraph width measures do indeed exist. Besides some simplistic examples, such as the measure that counts the number of vertices in the input graph, there is the treewidth of the underlying undirected graph. In the latter case we can apply the rich theory of (undirected) graphs of bounded treewidth, but we would not get anything substantially new for digraphs. As such, we are interested in digraph width measures that are *incomparable* to undirected treewidth.

---
[3] Kenny-width [9] is a different measure than Kelly-width [11].

Our second question is what properties can an algorithmically useful digraph width measure be expected to satisfy. In particular, can we expect any such properties typical for undirected width measures also in the directed case? An important feature of treewidth is that it allows a cops-and-robber game characterization. In fact, several digraph width measures such as DAG-width [2, 14], Kelly-width [11], and DAG-depth [9] admit some variants of a *cops-and-robber game* characterization. While there is no formal definition of a cops-and-robber game-based width measure, all versions of the cops-and-robber game that have been considered share a basic property: shrinking an induced (directed) path does not help the robber. To capture this phenomenon formally, we introduce the notion of a directed topological minor in Section 5. Essentially what we show is that a directed width measure that is "cops-and-robber-game-based" must be closed under directed topological minors. We note that there exist algorithmically useful measures other than undirected treewidth – digraph clique-width [6] and bi-rank-width [13] – which are not monotone even under taking subdigraphs.

Our main result (Theorem 6.5) then states that an algorithmically useful digraph width measure that is closed under directed topological minors cannot be substantially different from the treewidth of the underlying undirected graph. This implies that algorithmically useful digraph width measures different from treewidth of the underlying undirected graph cannot be based on a cops-and-robber game. One can ask whether an even stronger claim is true, namely whether closure under just taking subdigraphs is enough to refute the existence of algorithmically useful digraph width measures different from treewidth. We will show that this is not true by giving an explicit example in Theorem 6.6. Another interesting example is then given in Theorem 6.8.

The paper is organized in four parts. In Section 3, we formally establish and discuss the (above outlined) properties an algorithmically useful digraph width measure should have. In Section 4, we begin with the technical prerequisites for our main results. Briefly sketching, we will show that the structure of hard $MSO_1$-definable graph problems is as rich for planar graphs of degree at most 3 as for general graphs. In Section 5, we will introduce the notion of a directed topological minor. We will discuss its properties and consider complexity issues. In particular, we will show that it is hard to decide for a fixed (small) digraph whether it is a directed topological minor of a given digraph. In the last section, Section 6, we prove our main results which have already been outlined above.

Our proofs are based on some advanced techniques that are well known in undirected structural graph theory, but we apply them in a novel setting of digraph width measures. Due to lack of space, all the supplementary proofs can be found in the Appendix.

## 2 Definitions and notation

The graphs (both undirected and directed) that we consider in this paper are *simple* in that they do not contain loops. Given a graph $G$, we let $V(G)$ denote its vertex set and $E(G)$ denote its edge set, if $G$ is undirected. If $G$ is directed, we let $A(G)$ denote its arc set. The symbol $G$ usually denotes undirected graphs and $D$ denotes directed graphs. The arcs of a digraph $D$ are ordered pairs $(u,v) \in A(D)$ for $u \neq v$, where $u$ is an *in-neighbor* of $v$ and $v$ is an *out-neighbor* of $u$. Given a directed graph $D$, the *underlying undirected graph* $U(D)$ of $D$ is an undirected graph on the vertex set $V(D)$; and $\{u,v\}$ is an edge of $U(D)$ if and only if $(u,v) \in A(D)$ or $(v,u) \in A(D)$. A digraph $D$ is an *orientation* of an undirected graph $G$ if $U(D) = G$.

For a vertex pair $u, v$ of a digraph $D$, a sequence $P = (u = x_0, \ldots, x_r = v)$ is called *directed* $(u,v)$-*path* of length $r > 0$ in $D$ if the vertices $x_0, \ldots, x_r$ are pairwise distinct and $(x_i, x_{i+1}) \in A(G)$ for every $0 \leq i < r$. We also write $u \to_D^+ v$ if there exists a directed $(u,v)$-path in $D$, and $u \to_D^* v$ if either $u \to_D^+ v$ or $u = v$. A *directed cycle* is defined analogously with the modification that $x_0 = x_r$. A digraph $D$ is *acyclic* (a DAG) if $D$ contains no directed cycle.

A parameterized problem $Q$ is a subset of $\Sigma \times \mathbb{N}_0$, where $\Sigma$ is a finite alphabet. A parameterized problem $Q$ is said to be *fixed-parameter tractable* if there is an algorithm that given $(x, k) \in \Sigma \times \mathbb{N}_0$ decides whether $(x, k)$ is a yes-instance of $Q$ or not in time $f(k) \cdot p(|x|)$ time where $f$ is some computable function of $k$ alone, $p$ is a polynomial and $|x|$ is the size measure of the input. The class FPT denotes the class of parameterized problems that are fixed-parameter tractable. The class XP is the class of parameterized problems that admit algorithms with a run-time of $O(|x|^{f(k)})$ for some computable $f$.

Monadic second-order (MSO in short) logic is a language particularly suited for description of problems on "tree-like structured" graphs. For instance, the celebrated result of Courcelle [4], and of Arnborg, Lagergren and Seese [1], states that all $MSO_2$ definable graph problems have linear-time FPT algorithms when parameterized by the undirected treewidth. The expressive power of $MSO_2$ is very strong, as it includes many natural graph problems. In this paper we are, however, interested primarily in another logical dialect commonly abbreviated as $MSO_1$, whose expressive power is noticeably weaker than that of $MSO_2$. The weaker expressive power is not a handicap but an advantage for our paper since we are going to use it in negative results. Similarly to the previous, $MSO_1$ definable graph problems have FPT algorithms when parameterized by clique-width [5] and, consequently, by rank-width.

**Definition 2.1.** The language of $MSO_1$ contains the logical expressions that are built from the following elements:

- variables for elements (vertices) and their sets, and the predicate $x \in X$,
- the predicate $adj(u, v)$ with $u$ and $v$ vertex variables,
- equality for variables, the connectives $\wedge, \vee, \neg, \to$ and the quantifiers $\forall, \exists$.

*Example 2.2.* For an undirected graph to have the 3-colorability property is an $MSO_1$-expression:

$$\exists V_1, V_2, V_3 \left[ \forall v \, (v \in V_1 \vee v \in V_2 \vee v \in V_3) \wedge \bigwedge_{i=1,2,3} \forall v, w \, (v \notin V_i \vee w \notin V_i \vee \neg adj(v,w)) \right]$$

A decision graph property $\mathcal{P}$ is $MSO_1$ *definable* if there exists an $MSO_1$ formula $\phi$ such that $\mathcal{P}$ holds for any graph $G$ if, and only if, $G \models \phi$, i.e., $\phi$ is true on the model $G$. $MSO_1$ is analogously used for digraphs and their properties, where the predicate $arc(u, v)$ is used instead of $adj(u, v)$.

## 3 Desirable digraph width measures

A *digraph width measure* is a function $\delta$ that assigns each digraph a non-negative integer. To stay reasonable, we expect that infinitely many non-isomorphic digraphs are of bounded width. We consider what properties a width measure is expected to have. Importantly, one must be able to solve a rich class of problems on digraphs of bounded width. But what does "rich" mean?

On one hand, looking at existing algorithmic results in the undirected case, it appears that a *good balance* between the richness of the class of problems we capture and the possibility of positive general algorithmic results is achieved by the class of $MSO_1$ expressible problems (Definition 2.1). On the other hand, if we consider any logical language $\mathcal{L}$ over digraphs that is powerful enough to deal with sets of singletons (i.e. of monadic second order) and that can identify the adjacent pairs of vertices of the digraph, then we see $\mathcal{L}$ can naturally interpret also the $MSO_1$ logic of the underlying digraph. Hence the following specification appears to be the most natural common denominator in our context:

**Definition 3.1.** A digraph width measure $\delta$ is *powerful* if, for every $MSO_1$ definable decision property $\mathcal{P}$, there is an XP algorithm deciding $\mathcal{P}$ on all digraphs $D$ with respect to the parameter $\delta(D)$.

The traditional measures treewidth, branchwidth, clique-width, and more recent rank-width, are all powerful [4, 5] for undirected graphs. For directed graphs, unfortunately, exactly the opposite holds. The width measures suggested in recent years as possible extensions of treewidth – including directed treewidth [12], D-width [20], DAG-width [14, 2], and Kelly-width [11] – all are not powerful.

Another concern is about "non-similarity" of our directed measure $\delta$ to the traditional treewidth of the underlying undirected graph; we actually want to obtain and study new measures that significantly differ from treewidth, in the negative sense of the following Definition 3.2. This makes sense because any measure $\delta$ which bounds the treewidth of the underlying graph would automatically be powerful but wouldn't help solve any more problem instances than we already can with traditional undirected measures.

**Definition 3.2.** A digraph width measure $\delta$ is called *treewidth-bounding* if there exists a computable function $b$ such that, for every digraph $D$, $\delta(D) \leq k$ implies that the treewidth of $U(D)$ is at most $b(k)$.

To briefly outline the current state, we focus in the rest of this section on two of the treewidth-like directed measures which seem to attract most attention nowadays – DAG-width [14, 2] and Kelly-width [11]; and on another two not-much-known but significantly more successful (in the algorithmic sense) measures – directed clique-width [6] and bi-rank-width [13]. Of course, none of these measures is treewidth-bounding.

Since the definitions of DAG-width and Kelly-width are not short, we skip them here and refer to [14, 2, 11] instead. Both DAG- and Kelly-width share some common properties important for us:
– Acyclic digraphs (DAGs) have width 0 and 1, respectively.
– If we replace each edge of a graph of treewidth $k$ by a pair of opposite arcs, then the resulting digraph has DAG-width $k$ and Kelly-width $k + 1$.
– Both of the measures are characterized by certain cops-and-robber games.

**Proposition 3.3.** *Unless* P = NP, *DAG-width and Kelly-width are not powerful.*

On the other hand, there is the clique-width measure [6] which, although originally considered undirected, readily extends from graphs to digraphs. The *clique-width* of a graph $G$ is the smallest integer $k$ such that $G$ is the value of a $k$-expression. A *k-expression* is an algebraic expression with the following four operations on vertex-labeled graphs using $k$ labels: create a new vertex with label $i$; take the disjoint union of two labeled graphs; add all edges (arcs in the directed variant) between the vertices of label $i$ and label $j$; and relabel all the vertices with label $i$ to have label $j$.

Another noticeable directed measure is *bi-rank-width* [13], which is strongly related to clique-width in the sense that one is bounded on a digraph class iff the other one is. Due to restricted space we, however, only refer to [13] or [10] for its definition and properties.

**Proposition 3.4 (Courcelle, Makowsky, and Rotics [5]).** *Directed clique-width, and consequently bi-rank-width, are powerful measures.*

For a better understanding of the situation, we note one important but elusive fact: Bounding the undirected clique-width or rank-width of the underlying undirected graph does not generally help solve directed graph problems. In particular, undirected clique-width or rank-width are *not* powerful digraph measures. This is in a sharp contrast to the situation with treewidth where bounding the treewidth of the underlying undirected graph allows all the algorithmic machinery to work also on digraphs. A brief informal explanation of this antagonism lies in the facts that a "bag" in a tree decomposition has bounded size and so there is only a bounded number of possible orientations of the edges in it, while a single edge-addition operation in a clique-width expression creates a bipartite clique of an arbitrary size which admits an unbounded number of possible orientations. Consequently, there is no simple general relation between undirected measures and their directed generalizations.

After all, comparing Propositions 3.3 and 3.4, we clearly see the advantages of directed clique-width. There is, however, also the other side. Clique-width and bi-rank-width do not possess the nice structural properties common to the various treewidth-like measures, such as being subgraph- or contraction-monotone. This is due to symmetric orientations of complete graphs all having clique-width 2 while their subdigraphs include all digraphs, even those with arbitrarily high clique-width. This seems to be a drawback and a possible reason why clique-width- and rank-width-like measures are not so widely accepted.

The natural question now is; can we take *the better of each of the two worlds*? In our search for the answer, we will not study specific digraph width measures but focus on general properties of possible width measures. The main result of this paper, Theorem 6.5, then answers this question negatively: One *cannot* have a "nice" digraph width measure which is powerful, not treewidth-bounding and, at the same time, monotone under taking subgraphs and directed topological minors (see in Section 5). This strong and conceptually new result holds modulo technical assumptions which prevent our digraph width measures from "cheating", such as in Theorems 6.6 and 6.8.

## 4  Hard $MSO_1$ problems for $\{1,3\}$-regular planar graphs

The purpose of this section is to show that one can find many $MSO_1$ definable problems which are hard even on very restricted undirected graph classes such as on $\{1,3\}$-regular planar graphs. For a set $S$ of natural numbers, an *$S$-regular graph* is a graph having every vertex degree in $S$. This technical result will be essentially used in the proof of Theorem 4.1.

**Theorem 4.1.** *For any simple undirected graph $H$ and every $MSO_1$ formula $\varphi$, there exist a $\{1,3\}$-regular planar graph $G$ and an $MSO_1$ formula $\psi$, such that*

a) $H \models \varphi \iff G \models \psi$, *and*
b) *for every subdivision $G_1$ of $G$, it is $G_1 \models \psi \iff G \models \psi$.*
c) *Moreover, $\psi$ depends only on $\varphi$, $|\psi| = \mathcal{O}(|\varphi|)$, and both $G$ and $\psi$ are computable in polynomial time from $H$ and $\varphi$, respectively.*

Our main tool for the proof of this theorem is the classical interpretability of logic theories [15]. To describe its simplified idea, assume that two classes of *relational structures* $\mathscr{K}$ and $\mathscr{L}$ are given. The basic idea of an *interpretation $I$* of the theory $\text{Th}_{\text{MSO}}(\mathscr{K})$ into $\text{Th}_{\text{MSO}}(\mathscr{L})$ is to transform MSO formulas $\chi$ over $\mathscr{K}$ into MSO formulas $\chi^I$ over $\mathscr{L}$, and conversely structures $G \in \mathscr{L}$ into structures $G^I \in \mathscr{K}$, in such a way that "truth is preserved". Formal details are in the Appendix, and a brief illustration follows:

$$\begin{array}{ccc} \chi \in \text{MSO over } \mathscr{K} & \xrightarrow{I} & \chi^I \in \text{MSO over } \mathscr{L} \\ H \in \mathscr{K} & & G \in \mathscr{L} \\ \\ G^I \cong H & \xleftarrow{I} & G \end{array}$$

**Definition 4.2.** Let $\mathscr{K}$ and $\mathscr{L}$ be classes of relational structures. Theory $\text{Th}_{\text{MSO}}(\mathscr{K})$ is *interpretable* in theory $\text{Th}_{\text{MSO}}(\mathscr{L})$ if there exists an interpretation $I$ as above, such that the following two conditions are satisfied:

– For every structure $H \in \mathscr{K}$, there is $G \in \mathscr{L}$ such that $G^I \cong H$, and
– for every $G \in \mathscr{L}$, the structure $G^I$ is isomorphic to some structure of $\mathscr{K}$.

Furthermore, $\text{Th}_{\text{MSO}}(\mathscr{K})$ is *efficiently interpretable* in $\text{Th}_{\text{MSO}}(\mathscr{L})$ if the translation of each $\chi$ into $\chi^I$ is computable in polynomial time, and also the structure $G \in \mathscr{L}$ such that $G^I \cong H$ can be computed from any $H \in \mathscr{K}$ in polynomial time.

Since interpretability is clearly a transitive concept, we prove Theorem 4.1 in the following sequence of three relatively easy claims.

**Lemma 4.3.** *The $MSO_1$ theory of all simple undirected graphs is efficiently interpretable in the $MSO_1$ theory of simple planar graphs.*

**Lemma 4.4.** *The $MSO_1$ theory of all simple undirected graphs is efficiently interpretable in the $MSO_1$ theory of simple $\{1,3\}$-regular graphs. Moreover, this interpretation preserves planarity.*

**Lemma 4.5.** *For every $MSO_1$ formula $\varphi$ there exists an efficiently computable $MSO_1$ formula $\varphi_1$ such that the following holds: For every $\{1,3\}$-regular graph $G$ and every subdivision $G_1$ of $G$, it is $G_1 \models \varphi_1$ if and only if $G \models \varphi$.*

*Proof (of Theorem 4.1).* We apply the chain of interpretations $I_1, I_2, I_3$ from Lemmas 4.3, 4.4, and 4.5 in this order to the formula $\varphi$, obtaining the resulting formula $\psi \equiv ((\varphi^{I_1})^{I_2})^{I_3}$. We also construct, following the constructive proofs of the aforementioned lemmas, a graph $G$ such that $H \cong (G^{I_2})^{I_1}$ (notice the reverse order of the interpretations) Then part a) follows from Definition 4.2, and specially part b) from Lemma 4.5. Lastly, c) is true since the interpretations $I_1, I_2, I_3$ are efficient (meaning all computable in polynomial time). □

## 5 Directed topological minors

Many "reasonable" measures (treewidth being the prime example [17]) for undirected graphs are *monotone* under taking minors. In other words, the measure of a minor is not larger than the measure of the graph itself. Graph $H$ is a *minor* of a graph $G$ if it can be obtained by a sequence of applications of three operations: vertex deletion, edge deletion and edge contraction. (See e.g. [7].) It is therefore only natural to expect that a "nice" digraph measure should also be closed under some notion of a directed minor. However, there is currently no widely agreed definition of a directed minor. One published, but perhaps too restrictive on subdivisions, notion is the *butterfly minor* [12].

In order to be as general as possible, we will consider *directed topological minors*. The topological minor for undirected graphs is defined similarly to the usual minor, but an edge contraction can only be applied to an edge $e = \{u, v\}$ such that $u$ or $v$ has exactly two neighbors. In other words, graph $H$ is a topological minor of $G$ iff a subdivision of $H$ is isomorphic to a subgraph of $G$. To define directed topological minors, we first need a notion of *arc contraction* for digraphs:

**Definition 5.1.** Let $D$ be a digraph and $a = (x, y) \in A(D)$ be an arc. Then $D/a = ((V \setminus \{x, y\}) \cup \{v_a\}, A')$ is the digraph obtained by *contracting arc* $a$, where $v_a$ is a new vertex, and $(u, v) \in A'$ iff one of the following holds:

$$(u, v) \in A(D \setminus \{x, y\})$$
$$v = v_a \text{ and } (u, x) \in A \text{ or } (u, y) \in A$$
$$u = v_a \text{ and } (x, v) \in A \text{ or } (y, v) \in A$$

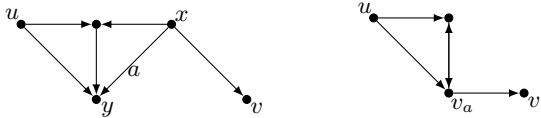

**Fig. 1.** Arc contraction: digraphs $D$ (left) and $D/a$.

See Fig. 1 for an example of a contraction. Note that contraction never produces arcs of the form $(x,x)$ (loops). Also the result of a contraction does not depend on the orientation of the contracted arc. Finally, we treat contracting a pair of arcs $(x,y)$ and $(y,x)$ as a contraction of a single bidirectional arc.

An important point of decision when defining a minor is; which arcs do we allow to contract? In the case of undirected graph minors, any edge can be contracted. For topological minors, only edges with at least one endvertex of degree two can be contracted. However, the situation is not so obvious in the case of digraphs. Look again at Figure 1. If we contract the arc $a$, we actually introduce a new directed path $u \to^+ v$, whereas in undirected graphs no new (undirected) path is ever created by the edge contraction.

On the other hand, simply never introducing a new directed path is not a good strategy either (for reasons which are explained in Remark 6.7). We denote by $V_3(D) \subseteq V(D)$ the subset of vertices having at least three neighbors in $D$. The middle ground we choose is to prohibit introducing a new directed path between any two vertices from $V_3(D)$. This is reflected in the following definition:

**Definition 5.2.** Let $D$ be a digraph and $a \in A(D)$. An arc $a = (u,v) \in A(D)$ is *2-contractible* if

- $u$ or $v$ has exactly two neighbors, and
- $(v,u) \in A(D)$ or there is no pair of vertices $x,y \in V_3(D)$ such that $x \to^*_{(D-a)} v$ and $u \to^*_{(D-a)} y$.

A digraph $H$ is a *directed topological minor* of $D$ if there exist digraphs $D_0, \ldots, D_r \cong H$ such that $D_0$ is a subgraph of $D$, and for all $0 \leq i \leq r-1$, $D_{i+1}$ is obtained from $D_i$ by contracting a 2-contractible arc.

**Proposition 5.3.** *Let $D$ be a digraph and $D'$ be a digraph obtained from $D$ by a sequence of vertex deletions, arc deletions and contractions of 2-contractible arcs. Then $D'$ is a directed topological minor of $D$.*

A useful notion in reasoning about directed topological minors is that of a 2-path. Let $D$ be a digraph and $P = (x_0, \ldots, x_k)$ a sequence of vertices of $D$. Then $P$ is a *2-path* (of length $k$) in $D$ if $P$ is a path in the underlying graph $U(D)$ and all internal vertices $x_i$ for $0 < i < k$ have exactly two neighbors in whole $D$. Obviously not every 2-path is a directed path. The following lemma explains the close relationship between 2-paths and directed topological minors.

**Lemma 5.4.** *Let $D$ be a digraph and $S = (x_0, \ldots, x_k)$ a 2-path of length $k > 2$ in $D$. Then there exists a sequence of 2-contractions of arcs of $S$ in $D$ turning $S$ into a 2-path of length two (or even of length one if $S$ was a directed path).*

The obvious question is whether the known digraph measures are closed under taking directed topological minors. The answer is given by the following proposition:

**Proposition 5.5.** *DAG-width and Kelly-width are monotone under taking directed topological minors unless the width is 0 or 1, respectively. Directed clique-width and bi-rank-width are not.*

In the second part of this section, we consider the complexity of deciding whether a given digraph is a directed topological minor of another digraph. We show that this problem is hard by giving a reduction from the 2-LINKAGE problem, which is the following problem. Let $D$ be a digraph and let $s_1, s_2, t_1, t_2$ be pairwise different vertices of $D$. A *2-linkage* for $\{(s_1, t_1), (s_2, t_2)\}$ is a pair $(P_1, P_2)$ of vertex-disjoint directed paths where $P_i$ is a $(s_i, t_i)$-path in $D$ for $i \in \{1, 2\}$.

**Proposition 5.6 ([8]).** *The* 2-LINKAGE *problem, given a digraph $D$ and $\{(s_1, t_1), (s_2, t_2)\}$ where $s_1, s_2, t_1, t_2$ are pairwise different vertices of $D$, to decide whether $D$ has a 2-linkage for $\{(s_1, t_1), (s_2, t_2)\}$, is* NP-*complete.*

**Theorem 5.7.** *There exists a digraph $H$ such that the problem, given a digraph $D$, to decide whether $H$ is directed topological minor of $D$, is* NP-*complete.*

The complexity result of Theorem 5.7 shows that it is already difficult for relatively simple digraphs to decide whether they are directed topological minor of some given digraph. I.e. the "directed topological minor" decision problem is not fixed-parameter tractable with the number of vertices of the minor as a parameter.

A natural question is to ask how the "directed topological minor" problem behaves on restricted input digraphs. It has been shown that the generalization of the 2-LINKAGE problem to arbitrary numbers of given pairs, that is called the LINKAGE problem (given a digraph $D$ and pairs of pairwise different vertices, decide whether the pairs can be joined by vertex-disjoint directed paths) is NP-complete on acyclic digraphs [21]. It is not difficult to see that the proof of Theorem 5.7 can be extended to prove the next result. In particular, if the input digraph $D$ is acyclic, all construction steps yield again acyclic digraphs.

**Theorem 5.8.** *The problem, given two acyclic digraphs $D$ and $H$, to decide whether $H$ is directed topological minor of $D$, is* NP-*complete.*

## 6 An (almost) optimal closure property result for digraph width measures

In this section we finally prove some "almost optimal" negative answers to the intrusive questions raised in the Introduction and at the end of Section 3. To recapitulate, we have asked whether it is possible to define a digraph width measure that is closed under some notion of a directed minor and that is still powerful (such as ordinary treewidth in the undirected sense). Notions of a minor and of possible directed minors have been surveyed in Section 5. We also recall the property of being treewidth-bounding (which we want to avoid) from Definition 3.2.

Besides the aforementioned several properties we suggest one more technical property that a desired nice directed width measure should posses to avoid "cheating" such as in the example of Theorem 6.8. Informally, we do not want to allow the measure to keep "computationally excessive" information in the orientation of edges:

**Definition 6.1.** A digraph width measure $\delta$ is *efficiently orientable* if there exist a computable function $h$, and a polynomial-time computable function $r : \mathscr{G} \to \mathscr{D}$ (from the class of all graphs to that of digraphs), such that for every undirected graph $G \in \mathscr{G}$, it is $U(r(G)) = G$ and

$$\delta(r(G)) \leq h(\min\{\delta(D) : D \text{ a digraph s.t. } U(D) = G\}).$$

**Proposition 6.2.** *DAG-width, Kelly-width, and digraph clique-width are all efficiently orientable.*

Our main proof also relies on some deep ingredients from Graph Minors:

**Theorem 6.3 ([19]).** *Let $H$ be a planar undirected graph. There exists a number $n_H$ such that for every undirected graph $G$ of treewidth at least $n_H$, $H$ is a minor of $G$.*

**Proposition 6.4 (folklore).** *If $H$ is a minor of $G$ and the maximum degree of $H$ is three, then $H$ is a topological minor of $G$.*

With all the ingredients at hand, we state and prove our main result:

**Theorem 6.5.** *Let $\delta$ be a digraph width measure with the following properties*

*a) $\delta$ is not treewidth-bounding;*
*b) $\delta$ is monotone under taking directed topological minors;*
*c) $\delta$ is efficiently orientable.*

*Then $P = NP$, or $\delta$ is not powerful.*

*Proof.* We assume that $\delta$ is powerful, and show a polynomial-time algorithm for solving any $MSO_1$ definable property $\varphi$ of undirected graphs, applying Theorem 4.1. Since, e.g. Example 2.2, there are NP-hard such properties, it would follow that $P = NP$.

Let $k$ be a suitable constant depending on $\delta$, as specified below. Let $\psi$ be the formula constructed from our $MSO_1$ formula $\varphi$, and let $G$ be the $\{1,3\}$-regular graph constructed from an arbitrary undirected graph $H$, both as in Theorem 4.1. Let moreover $G_1$ be the 1-subdivision of $G$ (i.e. replacing every edge of $G$ with a path of length two). We claim that, under assumptions a),b), there exists an orientation $D$ of $G_1$ such that $\delta(D) \leq k$.

We postpone the proof of this claim, and show its implications first. By c) Definition 6.1, we can efficiently construct an orientation $D_1$ of $G_1$ such that $\delta(D_1) \leq h(k)$ (a constant). Let $\psi_1$ be the (directed) $MSO_1$ formula obtained from $\psi$ by replacing $adj(u,v)$ with $(arc(u,v) \vee arc(v,u))$. Then, by Theorem 4.1, $H \models \varphi$ iff $D_1 \models \psi_1$, and hence we have got a polynomial reduction of the problems $H \models \varphi$ onto $D_1 \models \psi_1$. Since $\delta$ is assumed powerful, the latter problem can be solved by an XP algorithm wrt. constant parameter $h(k)$, i.e. in polynomial time.

Now we return back to our claim. Since a) $\delta$ is not treewidth-bounding, there is an integer $k \geq 0$ such that the class of all $U(D)$, where $D$ is a digraph of $\delta(D) \leq k$, has unbounded treewidth. So by Theorem 6.3, there exists $D_0$ such that $\delta(D_0) \leq k$ and $U(D_0)$ contains a $G_1$-minor. Since the maximum degree of $G_1$ is three, by Proposition 6.4 we have that some subdivision $G_2$ of $G_1$ is a subgraph of $U(D_0)$, or that some digraph $D_2$, $U(D_2) = G_2$, is a subdigraph of $D_0$. Then b) $\delta(D_2) \leq k$. Moreover, by Lemma 5.4 there exists a directed minor $D_3$ of $D_2$ such that $U(D_3) \cong G_1$, and $\delta(D_3) \leq k$ by b), too. We are done. □

Due to Theorem 6.5, a powerful digraph width measure essentially "cannot be stronger" than ordinary undirected treewidth, unless P equals NP. Our result requires two assumptions about the width measure $\delta$ in consideration: $\delta$ should be closed under taking directed topological minors, and it should be efficiently orientable. An interesting question is whether these conditions are necessary, or, put differently, whether the result of Theorem 6.5 can be strengthened by weakening these assumptions.

We address this question in the remaining part of this section – we show that Theorem 6.5 is almost strongest possible in the following Theorems 6.6 and 6.8. Specifically, we show that if one relaxes either of these two conditions, then one can construct powerful measures which definitely do not "look nice" (and, as such, one would not like to include them among the desired directed measures). In the first round, we relax the condition of monotonicity under taking directed topological minors just to subdigraphs:

**Theorem 6.6.** *There exists a powerful digraph width measure $\delta$ with the properties:*

*a) $\delta$ is not treewidth-bounding;*
*b) $\delta$ is monotone under taking subdigraphs;*
*c) $\delta$ is efficiently orientable.*

*Remark 6.7.* Theorem 6.6 can be slightly strengthened by claiming that $\delta$ is even monotone under such contractions of arcs $a \in A(D)$ that create no new directed paths in $D/a$ (compare to Definition 5.2 and the butterfly minors). In this modification we let $\delta(D) = 1$ if $\text{dist}_{U(D)}(u,v) \geq g(2 \cdot |V_3(D)|)$ for all $u \neq v \in V_3(D)$, and all the 2-paths between $u,v \in V_3(D)$ have "alternately oriented" arcs. Then every contraction in such $D$ creates a new directed path.

In the second round, we take a closer look at the condition that $\delta$ is efficiently orientable. It is not unreasonable to assume a digraph width measure to be efficiently orientable since most known digraph measure are, e.g. Proposition 6.2. Furthermore, efficient orientability prevents digraph measures from "keeping excessive information" in the orientation of arcs, such as (Theorem 6.8) the information about 3-colorability of the underlying graph. To carefully explain our point, note that there is nothing specially interesting in solving 3-colorability on digraphs; this claim is to show that an "NP-completeness oracle" can be encoded in the orientation of arcs of a digraph in a way that it is even preserved under taking directed topological minors, and hence efficient orientability is a natural required property of a desirable directed measure.

**Theorem 6.8.** *There exists a digraph width measure $\delta$ such that*

*a) $\delta$ is not treewidth-bounding;*
*b) $\delta$ is monotone under taking directed topological minors;*
*c) for every $k \geq 1$, on any digraph $D$ with $\delta(D) \leq k$, one can decide in time $\mathcal{O}(3^k \cdot n^2)$ whether $U(D)$ is 3-colorable, and find a 3-coloring if it exists.*

## 7 Conclusions

The main result of this paper shows that an algorithmically useful digraph width measure that is substantially different from treewidth cannot be closed under taking directed topological minors. Since cops-and-robber games behave invariantly on directed topological minors, we can conclude that a digraph width measure that allows efficient decisions of $\text{MSO}_1$-definable digraph properties on classes of bounded width should not be definable using the "standard" cops-and-robber games. This gives more weight to the argument [9] that bi-rank-width [13] is the best (though not optimal) currently known candidate for a *good* digraph width measure.

Our main result also leaves room for other ways of overcoming the problems with the currently existing digraph width measures. We have asked for width measures that are powerful, i.e., all $\text{MSO}_1$-definable digraph properties are decidable in polynomial time on digraphs of bounded width. What happens if we relax this requirement? We can ask for more time, like subexponential running time, or we can ask for restricted classes of $\text{MSO}_1$-definable digraph properties. Currently, we are not aware of any noticeable progress in this direction. Another interesting direction for future research is a closer study of efficient orientability and directed topological minors.

Finally, we believe that the results and suggestions contained in our paper will lead to new ideas and research directions in the area of digraph width measures – an area that seems to be stuck at this moment.

# Appendix

### Additions to Section 3

*Additional remarks to Definition 3.1:* Notice that we do not associate our abstract width measure $\delta$ with any particular "decomposition" (which is a typical attribute of width measures). This is to stay on the more general side, and a construction of such a decomposition for the aforementioned measures is not a problem anyway.

We could have even imposed in Definition 3.1 a stronger requirement, that $\mathcal{P}$ is decidable by an FPT algorithm. The undirected measures would stay powerful even in this stronger sense, but our main goal is to give negative results for directed width measures and hence the weaker requirement (of the existence of an XP algorithm) will actually make our conclusions stronger (cf. Theorem 6.5).

*Proof (of Proposition 3.3).* We take any NP-complete $\text{MSO}_1$ definable property $\mathcal{P}$ of undirected graphs, such as the 3-colorability from Example 2.2, and define the digraph $\text{MSO}_1$ property $\mathcal{P}'$ by replacing every occurrence of the predicate $adj(x, y)$ with $\bigl(arc(x, y) \vee arc(y, x)\bigr)$. Clearly, $\mathcal{P}$ holds on a graph $G$ if and only if $\mathcal{P}'$ holds on any orientation $D$ of $G$. If DAG-width or Kelly-width were powerful, then by Definition 3.1 the property $\mathcal{P}'$ would be decidable on all DAGs (width 0 / 1) in polynomial time. Hence for any input graph $G$ we could decide $\mathcal{P}$ by constructing an acyclic orientation $D$ of $G$, and then answering $\mathcal{P}'$ on $D$ in polynomial time. By standard arguments this would mean that P = NP. □

### Additions to Section 4

*Details of an interpretation $I$ of the theory $\text{Th}_{\text{MSO}}(\mathcal{K})$ into $\text{Th}_{\text{MSO}}(\mathcal{L})$.*

First one chooses a formula $\alpha(x)$ intended to define in each structure $G \in \mathcal{L}$ a set of individuals (new domain) $G[\alpha] := \{a : a \in dom(G) \text{ and } G \models \alpha(a)\}$, where $dom(G)$ denotes the set of individuals (domain) of $G$. Then one chooses for each $s$-ary relational symbol $R$ from $\mathcal{K}$ a formula $\beta^R(x_1, \ldots, x_s)$, with the intended meaning to define a corresponding relation $G[\beta^R] := \{(a_1, \ldots, a_s) : a_1, \ldots, a_s \in dom(G) \text{ and } G \models \beta^R(a_1, \ldots, a_s)\}$. With the help of these formulas one can define for each structure $G \in \mathcal{L}$ the relational structure $G^I := \bigl(G[\alpha], G[\beta^R], \ldots\bigr)$ intended to correspond with structures in $\mathcal{K}$.

There is also a natural way to translate each formula $\chi$ (over $\mathcal{K}$) into a formula $\chi^I$ (over $\mathcal{L}$), by induction on the structure of formulas. The atomic formulas are substituted by corresponding chosen formulas (such as $\beta^R$) with the corresponding variables. Then one proceeds via induction as follows:

$$(\neg \chi)^I \mapsto \neg(\chi^I), \quad (\chi_1 \wedge \chi_2)^I \mapsto (\chi_1)^I \wedge (\chi_2)^I,$$
$$\bigl(\exists x\, \chi(x)\bigr)^I \mapsto \exists y\, \bigl(\alpha(y) \wedge \chi^I(y)\bigr), \quad \bigl(\exists X\, \chi(X)\bigr)^I \mapsto \exists Y\, \chi^I(Y).$$

We believe Lemma 4.3 is a known statement, but since we have not found an explicit reference, we present an illustrating proof for it (in the Appendix).

*Proof (of Lemma 4.3).* For start, we define the formula $deg_1(x) \equiv \forall y, z\, [(adj(x,y) \wedge adj(x,z)) \to y = z] \wedge \exists y\, adj(x, y)$ expressing that $x$ is of degree 1 (in the model $G$). Furthermore, the formula $con(u, v, X) \equiv \forall Y \subseteq X \exists y, z\, \bigl[(y \in Y \vee y = u) \wedge (z \notin Y \vee z = v) \wedge adj(y, z)\bigr]$ expresses the property that the subgraph induced by $X \cup \{u, v\}$ connects $u$ to $v$, and $mcon(u, v, X) \equiv con(u, v, X) \wedge \forall Y (Y \subsetneq X \to \neg con(u, v, Y))$ says that $X$ is a minimal connection (a path) between $u, v$.

Let $\mathscr{G}$ be the class of all simple graphs, and $\mathscr{P}$ the class of planar simple graphs. Following Definition 4.2, the formula $\alpha_1$ defining the domain (vertex set in this case) of a graph $G^I \cong H \in \mathscr{G}$ inside $G \in \mathscr{P}$ is given as

$$\alpha_1(v) \equiv \neg deg_1(v) \land \forall x(adj(x,v) \to \neg deg_1(x)). \tag{1}$$

The underlying idea is that we would like to use some vertices of a planar graph $G \in \mathscr{P}$ to model edge "crossings" of $H \cong G^I$, see in Figure 2. Hence we "mark" each of such supplementary vertices with a new neighbor of degree 1.

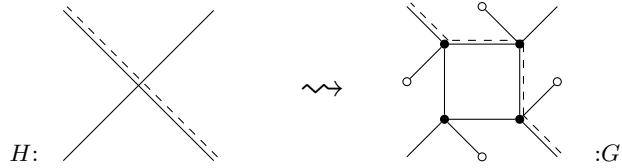

**Fig. 2.** The *crossing-gadget* modeling, in planar $G$, an edge crossing of $G^I \cong H$.

It remains to interpret the adjacency relation $\beta_1^{adj}(u,v)$ for $G^I$. Notice that the crossing gadget in Figure 2 uses a unique "marked" 4-cycle to model each crossing, and we can identify all such 4-cycles in $G$ with a formula $crgadg(C) \equiv \sigma \land (|C| = 4) \land \forall x \in C \neg \alpha_1(x)$ where $\sigma$ routinely describes the possible edge sets of a 4-cycle on a given set of vertices $C$. The shortcut $|.| = 4$ has an obvious implementation in MSO. Then we use

$$\begin{aligned}\beta_1^{adj}(u,v) \equiv \exists X \big[ &\forall x \in X\,(\neg \alpha_1(x)) \land mcon(u,v,X) \land \\ &\land \forall C\big((crgadg(C) \land X \cap C \neq \emptyset) \to |X \cap C| = 3\big)\big].\end{aligned} \tag{2}$$

The meaning of $\beta_1^{adj}(u,v)$ is that there exists a path $P$ between $u,v$ using only (besides $u,v$) marked internal vertices $X$, and such that $P$ intersects every crossing-gadget 4-cycle in exactly three vertices which ensure that $P$ is not "making a turn" at a crossing.

The last step in the proof is to verify the two conditions of Definition 4.2. While the second condition is trivially true since $\beta_1^{adj}$ is a symmetric binary relation, the first one requires an efficient algorithm constructing, for each $H \in \mathscr{K}$, a graph $G_H \in \mathscr{L}$ such that $G_H^I \cong H$. This is done as follows:

1. A "nice" drawing of $H$ in the plane is found (and fixed) such that no two edges cross more than once, no three edges cross in one point, and no edge passes through another vertex.
2. For every degree-1 vertex $w \in V(H)$, the unique edge $\{x,w\} \in E(H)$ is replaced with a path of length 3 on $\{x,w,w_1,w_2\}$ where $w_1,w_2$ are new vertices. (This is needed since degree-1 vertices have special meaning in the interpretation.)
3. Finally, every edge crossing is naturally replaced with a copy of the gadget from Figure 2. The resulting planar graph is named $G_H$.

It is routine to verify that $G_H^I \cong H$. □

*Proof (of Lemma 4.4).* Let $\mathscr{G}$ be the class of all simple graphs, and $\mathscr{R}$ denote the class of all simple $\{1,3\}$-regular graphs. We start the proof by showing a construction, for $H \in \mathscr{G}$, of the graph $G = G_H \in \mathscr{R}$ such that $G^I \cong H$ in the intended interpretation $I$.

For each vertex $v \in V(H)$ of degree $d$, we create a new vertex $r_v$ adjacent to two (three if $d = 0$) other new vertices of degree 1. If $d > 0$, then we also create a new bicolored (black–white) cycle $C_v$ of length $2d+2$ such that each its black vertex is adjacent to a new vertex of degree 1, one of the white vertices is adjacent to $r_v$, and the $d$ edges formerly incident with $v$ in $H$ are now one-to-one attached to the remaining $d$ white vertices of $C_v$. See Figure 3. The resulting graph is our $G_H$. Notice that if $H$ is planar, then the cyclic order of edges incident with each $C_v$ can be preserved, and so $G_H$ will also be planar.

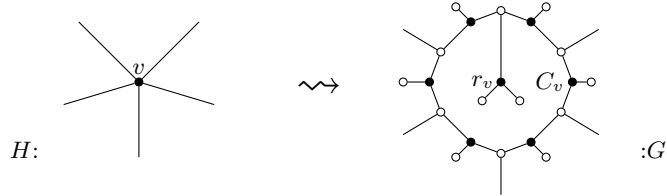

**Fig. 3.** The *vertex-gadget*, replacing vertices of $H$ with $\{1,3\}$-regular pieces in $G$.

The domain of $H$ can be identified within $G_H$ with the formula
$$\alpha_2(v) \equiv \exists x, y \, [\, x \neq y \wedge \mathrm{adj}(v,x) \wedge \mathrm{adj}(v,y) \wedge \tag{3}$$
$$\forall z \big((\mathrm{adj}(z,x) \vee \mathrm{adj}(z,y)) \to z = v\big) \,]$$
meaning simply that $v$ has (at least) two neighbors of degree 1.

Before interpreting the adjacency relation of $H$ in $G_H$, we have to identify the cycles $C_v$ from our construction. Notice that these are the only induced cycles of $G_H$ with the property that every second of their consecutive vertices has a neighbor of degree 1 (for instance, each edge of $G_H$ coming from an edge of $H$ has both ends with all neighbor of degree 3). In this sense we write
$$\varrho(U) \equiv \mathrm{cycle}(U) \wedge \forall x, y \in U \, \big[\mathrm{adj}(x,y) \to \tag{4}$$
$$\bigvee_{z=x,y} \exists w \, \forall t \, \big(\mathrm{adj}(z,w) \wedge (\mathrm{adj}(t,w) \to t = z)\big) \,\big]$$
where $\mathrm{cycle}(U)$ is a routine $\mathrm{MSO}_1$ predicate saying that $U$ induces a cycle in the graph, and finally,
$$\beta_2^{\mathrm{adj}}(u,w) \equiv \exists U, W \, \exists u_1, u_2 \in U, w_1, w_2 \in W \tag{5}$$
$$\varrho(U) \wedge \varrho(W) \wedge \mathrm{adj}(u_1, w_1) \wedge \mathrm{adj}(u_2, u) \wedge \mathrm{adj}(w_2, w).$$

We have finished description of the intended interpretation $I$, and it is now straightforward that $G_H^I \cong H$ for every $H \in \mathscr{G}$, cf. Definition 4.2. The proof is complete. □

*Proof (of Lemma 4.5).* An alternative view of the situation is that we are interpreting the $\mathrm{MSO}_1$ theory of $\{1,3\}$-regular graphs in the class of all their subdivisions. Hence we construct $\varphi_1$ as $\varphi^I$ in such an interpretation $I$: Since $G$ is $\{1,3\}$-regular, we simply identify the domain of $G$ (its vertex set) inside $G_1$ with $\alpha_3(v) \equiv \neg \mathrm{deg}_2(v)$ where $\mathrm{deg}_2(v)$ routinely expresses that $v$ is of degree two in $G_1$.

We moreover recall an $\mathrm{MSO}_1$ formula $\mathrm{con}(u,v,X)$ meaning that $u$ is connected to $v$ via the vertices of $X$ (in $G_1$). The adjacency relation of $G$ is then replaced with $\beta_3^{\mathrm{adj}}(u,v) \equiv \exists X \big(\mathrm{con}(u,v,X) \wedge \forall y \in X \, \mathrm{deg}_2(y)\big)$. Clearly, $G_1 \models \beta_3^{\mathrm{adj}}(u,v)$ if and only if $u$ and $v$ are connected with a path in $G_1$ created by subdividing an edge $\{u,v\}$ of $G$. The rest follows trivially. □

**Additions to Section 5**

*Proof (of Proposition 5.3).* From Definition 5.2 it follows that a 2-contractible arc cannot become non-2-contractible by vertex and arc deletions. Therefore we can first do the arc and vertex deletions by taking the corresponding subgraph $D''$ of $D$, and then do the contractions in the original order. Note that some of the arcs to be contracted may not be present in $D''$ because they have been removed by vertex deletions. □

*Proof (of Lemma 5.4).* If $S$ is a directed path in $D$, then any edge of $S$ is 2-contractible. Otherwise $S$ is not a directed path in $D$ and therefore there are two arcs $(x_i, x_{i+1}), (x_j, x_{j-1}) \in A(D)$ for some $i, j$ s.t. $0 \leq i < k$, $0 < j \leq k$ and $i+1 \neq j$. Then any other edge of $S$ is 2-contractible, as its contraction cannot produce a new directed path between any two vertices in $D$ with more than two neighbors. □

*Proof (of Proposition 5.5, sketch).* For DAG-width and Kelly-width this follows from the characterization by their corresponding cops-and-robber games: A 2-contraction does not create a new path between two vertices of degree greater than 2 (vertices of degree 2 have to be handled separately). The result for clique-width and bi-rank-width is a direct consequence of these two measures not being closed under taking subgraphs. □

*Proof (of Theorem 5.7).* Since the problem is clearly solvable in non-deterministic polynomial time, it remains to show that the problem is NP-hard. We reduce from the 2-LINKAGE problem. The proof goes in two steps. We first show that the 2-LINKAGE problem remains hard on digraphs where every vertex has at most three neighbors.

Let $D, \{(s_1, t_1), (s_2, t_2)\}$ be an instance of the 2-LINKAGE problem. Let $D'$ be a digraph such that $V(D') = V(D) \cup \{s'_1, s'_2, t'_1, t'_2\}$, where $s'_1, s'_2, t'_1, t'_2$ are new vertices, and $A(D') = A(D) \cup \{(s'_1, s_1), (s'_2, s_2), (t_1, t'_1), (t_2, t'_2)\}$. Obviously $D$ has a 2-linkage for $\{(s_1, t_1), (s_2, t_2)\}$ if and only if $D'$ has a 2-linkage for $\{(s'_1, t'_1), (s'_2, t'_2)\}$. Next, we modify large-degree vertices. We obtain digraph $D''$ from $D'$ by iteratively executing, for every vertex $x$ with $d_{D'}(x) \geq 4$, the following sequence of operations: delete $x$, introduce $d_{D'}(x)$ new vertices $x_1, \ldots, x_{d_{D'}(x)}$, add the arcs $(x_i, x_{i+1})$ for $1 \leq i < d_{D'}(x)$, assign the in-neighbors of $x$ as in-neighbors to the vertices $x_1, \ldots, x_j$ where $j = |N^{\text{in}}_{D'}(x)|$, and assign the out-neighbors of $x$ as out-neighbors to the remaining vertices. Observe that $x$ is replaced by a digraph with vertices of degree at most 3, and reachability is preserved. In particular, it holds that $D'$ has a 2-linkage for $\{(s'_1, t'_1), (s'_2, t'_2)\}$ if and only if $D''$ has a 2-linkage for $\{(s'_1, t'_1), (s'_2, t'_2)\}$. Since the construction of $D''$ requires only polynomial time, this shows that 2-LINKAGE is NP-complete on digraphs of maximum degree at most 3 due to Proposition 5.6.

For the second step of the proof, we continue with $D''$ and $\{(s'_1, t'_1), (s'_2, t'_2)\}$. We introduce three new vertices and make them in-neighbors of $s'_1$. Similarly, we make three new vertices out-neighbors of $t'_1$, and four new vertices become in-neighbors of $s'_2$ and another four new vertices become out-neighbors of $t'_2$. Let $D^*$ be the resulting digraph. It holds that $s'_1, s'_2, t'_1, t'_2$ are the only vertices of $D^*$ of degree more than 3. We want to show that $D^*$ has a 2-linkage for $\{(s'_1, t'_1), (s'_2, t'_2)\}$ if and only if the digraph $H$ depicted in Figure 7 is a directed topological minor of $D^*$. For the first implication, let $D^*$ have a 2-linkage for $\{(s'_1, t'_1), (s'_2, t'_2)\}$. Thus, there are a directed $(s'_1, t'_1)$-path $P_1$ and a directed $(s'_2, t'_2)$-path $P_2$ in $D^*$ that are vertex-disjoint. Hence, $D^*$ has a subgraph $F$ that contains the vertices of $P_1$ and $P_2$ and the fourteen new vertices for $D^*$, that are connected only to $s'_1, s'_2, t'_1, t'_2$. By Lemma 5.4 we can contract both $P_1$ and $P_2$ to a single arc each. Therefore $H$ is a directed topological minor of $D^*$. For the converse, let $H$ be a directed topological minor of $D^*$. Due to the definition, there is a subgraph $H'$ of $D^*$ such that $H$ is isomorphic to a digraph that is obtained from $H'$ by only contracting 2-contractible arcs. Since all vertices of $H$ have degree different from 2, all arcs of $H$ are either arcs of $H'$ or result of a contraction. In particular, the two vertices of degree 4 of $H$ correspond to $s'_1$ and $t'_1$ of $D^*$, and the two vertices of degree 5 of $H$ correspond to $s'_2$ and $t'_2$ of $D^*$. That is because these four vertices are the only vertices of $D^*$ of degree larger than 3. By definition of 2-contractible arc, a path between two vertices of a degree greater than two can be contracted to an arc only if it directed path between these two vertices. Hence, since $H$ is obtained from $H'$ by only contracting contractible arcs, there are a directed $(s'_1, t'_1)$-path $P_1$ and a directed $(s'_2, t'_2)$-path $P_2$ in $H'$ that are vertex-disjoint. Since $H'$ is a subgraph of $D^*$, $P_1$ and $P_2$ are directed paths in $D^*$, and therefore, $D^*$ has a 2-linkage for $\{(s'_1, t'_1), (s'_2, t'_2)\}$. This completes the proof of the theorem.    □

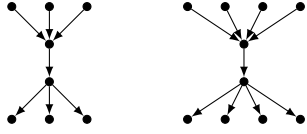

**Fig. 4.** Digraph $H$ from the proof of Theorem 5.7.

### Additions to Section 6

*Proof (of Proposition 6.2, sketch).* As noted in Section 3, DAG-width and Kelly-width attain their globally minimum values on DAGs. On the other hand, clique-width attains an optimal value on symmetric orientations of graphs (replacing each edge by a pair of opposite arcs).    □

*Proof (of Theorem 6.6).*

To show the existence of efficient algorithms for the $\delta$ we are going to define, we apply the following modified version of Courcelle's Theorem: *There exists a computable function $g$ such that for all digraphs $D$ and $MSO_2$ definable digraph properties $\varphi$, it can be decided in time $\mathcal{O}\big(g(|\varphi| + |V_3(D)|) \cdot |V(D)|\big)$ whether $D \models \varphi$.*

Recall that $V_3(D) \subseteq V(D)$ denotes the subset of those vertices having at least three neighbors in $D$. The original version of the above theorem gives actually a stronger result – using the quantifier depth (rank) of $\varphi$ instead of $|\varphi|$, and the treewidth of $U(D)$ instead of $|V_3(D)|$. Although Courcelle explicitly speaks about undirected $MSO_2$, the same result is clearly valid also for digraph $MSO_2$ (of course, with respect to the undirected treewidth), and for $MSO_1$.

We give an explicit definition of $\delta$. For an undirected graph $G$, we denote by $\text{dist}_G(u,v)$ the length of a shortest path between vertices $u$ and $v$ in $G$; if there is no $(u,v)$-path in $G$ then $\text{dist}_G(u,v) = \infty$. Let $g$ be the function as in Courcelle's theorem as stated previously. Without loss of generality, we can assume that $g$ is non-decreasing. For a digraph $D$, we define

$$\delta(D) = \begin{cases} 1, & \text{if } \text{dist}_{U(D)}(u,v) \geq g(2 \cdot |V_3(D)|) \text{ for all pairs } u,v \in V_3(D); \\ |V(D)|, & \text{otherwise}. \end{cases}$$

We first show that $\delta$ fulfills the claimed properties. For start, notice that $\delta$ does not depend on the orientation of edges; formally, $U(D_1) = U(D_2)$ readily implies $\delta(D_1) = \delta(D_2)$. Hence, c) $\delta$ is efficiently orientable in linear time. If we take any undirected graph $G$ (which can have arbitrarily large treewidth) and subdivide every edge of $G$ with $g(2 \cdot |V(G)|)$ vertices, then $\delta(D) = 1$ holds for every orientation $D$ of $G = U(D)$. Therefore, a) $\delta$ cannot be treewidth-bounding.

Third, concerning b), let $D$ be a digraph and let $F$ be a subdigraph of $D$. We have to show that $\delta(D) \geq \delta(F)$. This is clearly true if $\delta(D) = |V(D)|$, and so assume $\delta(D) = 1$. Take any vertex pair $u, v \in V_3(F) \subseteq V_3(D)$. Then by our assumption, $\mathrm{dist}_{U(D)}(u,v) \geq g(2 \cdot |V_3(D)|)$, and $g(2 \cdot |V_3(D)|) \geq g(2 \cdot |V_3(F)|)$ by assumed monotonicity of $g$. Hence $\mathrm{dist}_{U(F)}(u,v) \geq \mathrm{dist}_{U(D)}(u,v) \geq g(2 \cdot |V_3(F)|)$, and consequently $\delta(F) = 1 \leq \delta(D)$.

It remains to show that $\delta$ is powerful. Let $\varphi$ be an MSO$_1$ definable digraph property, and let $D$ be an input digraph. We simply apply Courcelle's theorem to decide $D \models \varphi$ and prove that this is an XP (even FPT) algorithm for the parameter $\delta(D)$. If $\delta(D) = |V(D)|$, then indeed $\mathcal{O}(g(|\varphi| + |V_3(D)|) \cdot |V(D)|) = \mathcal{O}(g(\delta(D)) \cdot |V(D)|)$ for every fixed $\varphi$. So assume $\delta(D) = 1$. If every component of $U(D)$ contains at most one cycle, then the treewidth of $D$ is at most two and the case follows trivially. Otherwise, some two vertices of $V_3(D)$ are connected by a path and so $|V(D)| \geq g(2 \cdot |V_3(D)|)$. Then the run-time bound of Courcelle's theorem gives $\mathcal{O}(g(|\varphi| + |V_3(D)|) \cdot |V(D)|) = \mathcal{O}(\max\{g(2|\varphi|), |V(D)|\} \cdot |V(D)|) = \mathcal{O}(|V(D)|^2)$ for fixed $\varphi$. □

*Proof (of Theorem 6.8).* We start by defining our digraph width measure $\delta$. For a digraph $D$,

$$\delta(D) = \begin{cases} 1 & \text{if the arcs of } D \text{ encode a 3-coloring of } U(D) \\ |V(D)| & \text{otherwise} \end{cases}$$

We say that a digraph $D$ *encodes a 3-coloring* if, for every directed path $s \to_D^+ t$ s.t. $s, t \in V_3(D)$, we have that $s$ has no in-neighbors (i.e. $s$ is a *source*) or $t$ has no out-neighbors (i.e. $t$ is a *sink*). The crucial property of this definition is that if $D$ encodes a 3-coloring, then $U(D)$ is 3-colorable. A *3-coloring* is a partition of vertices of $V(D)$ into three independent sets (an independent set consists of pairwise nonadjacent vertices). Let $S_1, S_2, S_3$ be a partition of $V_3(D)$ such that $S_1$ contains all sources, $S_3$ contains all sinks, and $S_2$ the remaining vertices. $S_1$ and $S_3$ are obviously independent (as they contain only source/sink vertices). $S_2$ is independent since $D$ encodes a 3-coloring. As vertices of $V(D) \setminus V_3(D)$ have at most two neighbors, we can use straightforward greedy algorithm to extend the $(S_1, S_2, S_3)$ partition of $V_3(D)$ to a partition of whole $V(D)$ into three independent sets. Therefore $U(D)$ is 3-colorable.

On the other hand let $S_1, S_2, S_3$ be a 3-coloring of a graph $G$. Then we create its orientation $D$ (such that $U(D) = G$) by directing the incident edges for each vertex in $v \in S_1$ away from $v$, and the incident edges for each vertex in $v \in S_3$ towards $v$. As $S_1, S_2, S_3$ form a 3-coloring, this process orients all edges and we get no conflicts. To see that $D$ encodes a 3-coloring it is enough to note that the directed paths in $D$ have length at most two.

We now have to show that $\delta$ fulfills the properties a,b,c. To prove that $\delta$ is not treewidth bounding let us consider $K_{n,n}$, the symmetric complete bipartite graph on $2n$ vertices. Let $D$ be a digraph created by orienting the edges of $K_{n,n}$ such that all sources are on one side of the bipartition. Then $\delta(D) = 1$, $U(D) = K_{n,n}$ and it is a well known fact that $\mathrm{tw}(K_{n,n}) = n - 1$. Since we can do this construction for every $n \geq 1$, $\delta$ is not treewidth bounding.

The next step is to show that $\delta$ is monotone under taking directed topological minors for every $k \geq 1$. Let $D$ be a digraph. If $\delta(D) \geq 2$, then $\delta(D) = |V(D)|$ and thus $\delta(H) \leq \delta(D)$ for every directed topological minor $H$ of $D$. Therefore let $\delta(D) = 1$. Then $D$ encodes a 3-coloring, and we have to show that every topological minor $H$ of $D$ also encodes a 3-coloring. First note that every subdigraph $D'$ of $D$ obviously encodes a 3-coloring, too. So let $a = (u,v)$ be a 2-contractible arc in $D$. Then $D' = D/a$ also encodes a 3-coloring, since this contraction does not create any new directed path between vertices in $V_3(D') \subseteq V_3(D) \cup \{v_a\}$.

To prove the third property, let $D$ be a digraph. If $\delta(D) = k \geq 2$, then $n = |V(D)| = k$. By trying all possible colorings we can test 3-colorability in time $\mathcal{O}(3^k \cdot n^2)$. On the other hand if $\delta(D) = 1$, then $D$ encodes a 3-coloring and we can compute a 3-coloring of $U(D)$ in time $\mathcal{O}(n^2)$, as outlined above. □